\documentclass[aps,prd,onecolumn,groupedaddress,showpacs,nofootinbib,amssymb]{revtex4-2}
\usepackage[dvips]{graphicx}
\usepackage{amssymb}
\usepackage{amsmath}
\usepackage{graphicx,,color}
\usepackage{amsfonts}
\usepackage{bm}
\usepackage{cancel}
\usepackage{comment}
\usepackage{floatflt}
\usepackage{slashed}
\usepackage{appendix}
\usepackage{array}
\usepackage{tabularx}

\newcommand\be{\begin{equation}}
\newcommand\ee{\end{equation}}

\allowdisplaybreaks[4]

\begin{document}

\tolerance=5000

\title{Imprints of a Second Order Electroweak Phase Transition on the Stochastic Gravitational Wave Background}
\author{V.K. Oikonomou,$^{1,2}$}\email{voikonomou@gapps.auth.gr;v.k.oikonomou1979@gmail.com}
\affiliation{$^{1)}$Department of Physics, Aristotle University of
Thessaloniki, Thessaloniki 54124, Greece \\
$^{2)}$ L.N. Gumilyov Eurasian National University - Astana,
010008, Kazakhstan}


\tolerance=5000

\begin{abstract}
In this work we shall study the impact of a second order
electroweak phase transition occurring at $\sim 150\,$GeV on the
energy spectrum of the stochastic gravitational background.
Specifically, we assume that the non-minimally coupled Higgs field
controls the inflationary era, we find the reheating temperature
for the Higgs inflationary model and we demonstrate that the Higgs
effective potential exhibits a very weak first order phase
transition. This weak first order phase transition is an
indication that the electroweak phase transition may not actually
proceed as a first order phase transition, but it will proceed as
a crossover or second order phase transition. This second order
phase transition proceeds with the Higgs field slow-rolling its
potential toward to its new minimum. This slow-rolling may deform
the radiation domination total equation of state, and the aim of
this work is to pinpoint the observational imprints of this total
equation of state deformation on the energy spectrum of the
primordial gravitational waves, that affects modes that enter the
horizon at temperatures $T\sim 150\,$GeV or lower.
\end{abstract}

\maketitle

\section*{Introduction}

Perceiving the primordial Universe remains a theoretical challenge
to date that remains to be understood. The most prominent theory
describing the primordial Universe is the inflationary paradigm,
which solves many theoretical shortcomings of the Big Bang
description. In the next decade, the inflationary era will be put
to severe testing, starting from the Simons observatory
\cite{SimonsObservatory:2019qwx} and hopefully the stage 4 cosmic
microwave background (CMB) experiments \cite{CMB-S4:2016ple} if
these commence to operate eventually. These CMB based experiments
have the ability to probe directly the inflationary era and verify
its existence if the B-mode of the inflationary era. Apart from
these direct probes of the inflationary era, the future stochastic
gravitational wave detectors
\cite{Hild:2010id,Baker:2019nia,Smith:2019wny,Crowder:2005nr,Smith:2016jqs,Seto:2001qf,Kawamura:2020pcg,Bull:2018lat,LISACosmologyWorkingGroup:2022jok}
can also probe inflationary gravitational waves and these are
quite promising for probing the inflationary era. On the antipode
of these observational tests, terrestrial experiments of particle
accelerators provided too little or nearly none information for
the physics beyond the Standard Model (SM). This can either mean
that there is none physics beyond the SM, a perspective rather
unlikely since many questions and problems of the SM are still
unanswered, or it means that we are technologically and
energetically far from probing beyond SM physics. In this line of
research, tri-linear Higgs couplings, which can probe directly the
electroweak symmetry breaking, will be probed 15-20 years from now
in CERN. This perspective is promising but not enough for the
complete understanding of the primordial Universe era and of the
physics beyond the SM.

Apart from these perspectives, even if inflation is finally
experimentally or observationally verified, there are many
challenges mainly theoretical to overcome. Firstly the
inflationary theory itself. Inflation can be realized by a
slow-rolling scalar field
\cite{inflation1,inflation2,inflation3,inflation4} or even
geometrically by some appropriate and well motivated modified
gravity theory \cite{reviews1,reviews2,reviews3,reviews4}. The
perspective of the scalar field realized inflation has some issues
to address consistently and a prominent question is related to the
many couplings of the inflaton to the SM model particles, needed
for reheating the cold and large Universe after the inflationary
era ends. Among many single scalar field models, the Higgs
inflation theory stands out
\cite{Bezrukov:2014bra,GarciaBellido:2011de,Bezrukov:2010jz,Bezrukov:2007ep,Mishra:2018dtg,Steinwachs:2013tr,Rubio:2018ogq,Kaiser:1994vs,Gundhi:2018wyz,CervantesCota:1995tz,Kamada:2012se,Schlogel:2014jea,Fuzfa:2013yba},
since the Higgs field plays the role of the inflaton itself and
thus the SM couplings of the inflaton are basically the Higgs
couplings to the SM particles. The Higgs inflation is compatible
with the latest Planck data \cite{Akrami:2018odb} and also very
well theoretically motivated. In this work we shall assume that
the Higgs field is responsible for the inflationary era. A single
minimally coupled Higgs inflationary theory is not a viable
option, but a non-minimally coupled single Higgs field can provide
a viable inflationary era. If we follow the evolution of the
Universe from the end of the inflationary era to deeply the
radiation domination era, after the Higgs field reheats the
Universe beyond the end of the inflationary era, the temperature
of the Universe increases until its maximum value is reached, the
so-called reheating temperature, after which the temperature of
the Universe drops. When the Universe's temperature is of the
order $\sim 150\,$GeV, the electroweak symmetry breaking occurs
which gives mass to the SM particles and also if it is a strong
first order transition, the necessary baryon asymmetry in the
Universe is generated. But the SM itself is not able to produce a
strong first order phase transition. Singlet extensions of the SM
are able to provide a strong first order phase transition, but one
needs an extra scalar field. If no such scalar field is used, the
electroweak phase transition is a second order
\cite{Kajantie:1996qd,Kajantie:1996mn} or crossover, see
\cite{Athron:2023xlk} for details. Adopting this line of
reasoning, in this work we shall assume that the Higgs field
experiences a second order phase transition at a temperature $\sim
150\,$GeV. This second order phase transition is materialized by a
slow-rolling of the Higgs field from its minimum at the origin
toward its new minimum. This slow-rolling of the Higgs field
during the radiation domination epoch can affect and finally
deform the total background equation of state (EoS), at the epoch
of the second order phase transition, and this EoS deformation may
have a direct imprint on the energy spectrum of the stochastic
gravitational waves, affecting frequencies equal or smaller than
the frequencies of the modes which enter the horizon during the
second order phase transition. As we show the effect can be
detectable by experiments like Litebird and this pattern of the
stochastic gravitational wave background is unique in the
literature.

This article is organized as follows: In section I we review the
formalism of non-minimally coupled scalar-tensor theories and how
to obtain the Einstein frame minimally coupled counterpart theory.
In section II we apply the formalism of section II to present the
essential features of the Higgs inflationary model and its
viability. In section III we explore the order of the reheating
temperature in the Higgs inflationary model. In section IV we
demonstrate explicitly the weakness of the Higgs-SM first order
phase transition and provide motivation for the second order
perspective of the electroweak phase transition. In section V we
present the effect of a second order electroweak phase transition
on the energy spectrum of the primordial gravitational waves.
Finally the conclusions follow at the end of the article.

\section{Non-minimal Coupled Scalar Field Theory in the Jordan and Einstein Frames: Formalism}

Since we will be interested in the Higgs inflation formalism, in
this section we provide a brief overview of non-minimally coupled
theories in the Jordan frame and their correspondence with their
Einstein frame counterparts, based on Refs.
\cite{Kaiser:1994vs,valerio,Faraoni:2013igs,Buck:2010sv,Faraoni:1998qx}.

We consider the Jordan frame action in the presence of a
non-minimally coupled scalar field action, and in the presence of
ordinary perfect fluid matter,
\begin{equation}\label{c1}
\mathcal{S}_J=\int d^4x\Big{[}f(\phi)R-\frac{1}{2}g^{\mu
\nu}\partial_{\mu}\phi\partial_{\nu}\phi-U(\phi)\Big{]}+S_m(g_{\mu
\nu},\psi_m)\, ,
\end{equation}
with $\psi_m$ denoting the perfect matter fluids, with their
pressure being $P$ and their energy density being $\epsilon$. Also
$g_{\mu \nu}$ denotes the Jordan frame metric. Note that the
minimal coupling choice for the above action corresponds to,
\begin{equation}\label{c2}
f(\phi)=\frac{1}{16 \pi G}=\frac{M_p^2}{2}\, ,
\end{equation}
with,
\begin{equation}\label{c3}
M_p=\frac{1}{\sqrt{8\pi G}}\, ,
\end{equation}
being the Jordan frame reduced Planck mass, $M_p=2.43\times
10^{18}$GeV, $G$ stands for the Jordan frame Newton's
gravitational constant. Upon performing the following conformal
transformation,
\begin{equation}\label{c4}
\tilde{g}_{\mu \nu}=\Omega^2g_{\mu \nu}\, ,
\end{equation}
or written differently,
\begin{equation}\label{c5}
\tilde{g}^{\mu \nu}=\Omega^{-2}g^{\mu \nu}
\end{equation}
one can obtain the Einstein frame gravitational action. Note that
the tilde denotes quantities considered in the Einstein frame. The
conformal factor $\Omega$ written in terms of the non-minimal
coupling function $f(\phi)$ reads
\cite{Kaiser:1994vs,Mishra:2018dtg,valerio},
\begin{equation}\label{c6}
\Omega^2=\frac{2}{M_p^2}f(\phi)\, .
\end{equation}
The Jordan frame quantities are transformed in the Einstein frame
as follows,
\begin{equation}\label{c7}
\sqrt{-g}=\Omega^{-4}\sqrt{-\tilde{g}}\, ,
\end{equation}
regarding the trace of the metric, while the Ricci scalar is
transformed as,
\begin{equation}\label{c8}
R=\Omega^2\left(\tilde{R}+6\tilde{\square}f-6\tilde{g}^{\mu
\nu}f_{\mu}f_{\nu}\right)\, ,
\end{equation}
while the d'Alembertian as follows,
\begin{equation}\label{c9}
\tilde{\square}f=\frac{1}{\sqrt{-\tilde{g}}}\tilde{\partial}_{\mu}\left(
\sqrt{-\tilde{g}}\tilde{g}^{\mu \nu}\partial_{\nu}f\right)\, .
\end{equation}
Hence, the term containing the non-minimal coupling has the
following transformation properties,
\begin{equation}\label{c10}
\int d^4x\sqrt{-g}f(\phi)R\to \int d^4x
\sqrt{-\tilde{g}}\frac{M_p^2}{2}\left(
\tilde{R}-6\left(\frac{1}{\Omega^2}\right)^2\tilde{g}^{\mu
\nu}\tilde{\partial}_{\mu}\Omega \tilde{\partial}_{\nu}\Omega
\right)\, ,
\end{equation}
while the Jordan frame kinetic term and the potential term have
the following transformation properties,
\begin{equation}\label{c11}
\int d^4 x \sqrt{-g} \Big{[}-\frac{1}{2}g^{\mu
\nu}\partial_{\mu}\phi \partial{\nu}-U(\phi)\Big{]} +S_m(g_{\mu
\nu},\psi_m)\to \int d^4 x
\sqrt{-\tilde{g}}\Big{[}-\frac{1}{2\Omega^2} \tilde{g}^{\mu
\nu}\tilde{\partial}_{\mu}\phi
\tilde{\partial}{\nu}\phi-\frac{U(\phi)}{\Omega^4(\varphi)}\Big{]}
+S_m(\Omega^2\tilde{g}_{\mu \nu},\psi_m)\, ,
\end{equation}
and hence, the Einstein frame action in terms of the conformal
factor $\Omega^2=\frac{2}{M_p^2}f$ reads,
\begin{equation}\label{c12}
\mathcal{S}_E=\int
d^4x\sqrt{-\tilde{g}}\Big{[}\frac{M_p^2}{2}\tilde{R}-\frac{\zeta
(\phi)}{2} \tilde{g}^{\mu \nu }\tilde{\partial}_{\mu}\phi
\tilde{\partial}_{\nu}\phi-V(\phi)\Big{]}+S_m(\Omega^{-2}\tilde{g}_{\mu
\nu},\psi_m)\, ,
\end{equation}
with,
\begin{equation}\label{c13}
V(\phi)=\frac{U(\phi)}{\Omega^4}\, ,
\end{equation}
and in addition we introduced the function $\zeta(\phi)$ which is
defined as,
\begin{equation}\label{c14}
\zeta
(\phi)=\frac{3}{2}M_p^2\frac{1}{f^2}\Big{(}\frac{df}{d\phi}\Big{)}^2+\frac{M_p^2}{2f}\,
.
\end{equation}
We can make the Einstein frame scalar field $\phi$ canonical, by
trying to make the function $\zeta (\phi)= 1$ by a new field
redefinition, so by rescaling,
\begin{equation}\label{c15}
\Big{(}\frac{d\varphi}{d \phi}\Big{)}^2 =\zeta(\phi)\, ,
\end{equation}
or similarly,
\begin{equation}\label{c16}
\frac{d \varphi}{d \phi}=
M_p\sqrt{\frac{1}{2f}+\frac{3}{2}\Big{(}\frac{f'}{f}\Big{)}^2}\, ,
\end{equation}
we get the Einstein frame action in terms of the canonical scalar
field $\varphi$,
\begin{equation}\label{c17}
\mathcal{S}_E=\int
d^4x\sqrt{-\tilde{g}}\Big{[}\frac{M_p^2}{2}\tilde{R}-\frac{1}{2}\tilde{g}^{\mu
\nu } \tilde{\partial}_{\mu}\varphi
\tilde{\partial}_{\nu}\varphi-V(\varphi)\Big{]}+S_m(\Omega^2\tilde{g}_{\mu
\nu},\psi_m)
\end{equation}
with,
\begin{equation}\label{c18}
V(\varphi)=\frac{U(\varphi)}{\Omega^4}=\frac{U(\varphi)}{4
M_p^4f^2}\, .
\end{equation}
One major difference between the Jordan frame and Einstein frame
is that the matter fluids in the Einstein frame are not perfect
fluids, thus the baryons do not follow free fall geodesics. This
can be seen in the action (\ref{c17}) where the matter action
contains the conformal factor coupled to the metric
$S_m(\Omega^2\tilde{g}_{\mu \nu},\psi_m)$. Specifically the energy
momentum tensor,
\begin{equation}\label{c19}
\tilde{T}_{\mu \nu}=-\frac{2}{\sqrt{-\tilde{g}}}\frac{\delta
L_m}{\delta \tilde{g}^{\mu \nu}}\, ,
\end{equation}
transforms from the Jordan to the Einstein frame in the following
way,
\begin{equation}\label{c20}
\tilde{T}_{\mu \nu}=\Omega^{-2}(\varphi)T_{\mu \nu}\, ,
\end{equation}
\begin{equation}\label{c21}
\tilde{T}^{\mu}_{\nu}=\Omega^{-4}(\varphi)T^{\mu}_{\nu}\, ,
\end{equation}
\begin{equation}\label{c22}
\tilde{T}^{\mu \nu}=\Omega^{-6}(\varphi)T^{\mu \nu}\, .
\end{equation}
The continuity equation of the Einstein frame energy momentum
tensor is,
\begin{equation}\label{c24}
\tilde{\partial}^{\mu}\tilde{T}_{\mu \nu}=-\frac{d}{d\varphi}[\ln
\Omega]\tilde{T}\tilde{\partial}_{\nu}\phi\, ,
\end{equation}
from which it is apparent that it does not describe perfect fluid
quantities. However, for the study of the inflationary theory of
the Higgs model, the fact that matter fluids in the Einstein frame
are not perfect fluids is not a problem, since matter fluids are
disregarded from the inflationary dynamics analysis. The
quantities of interest that are conformal invariant quantities are
the inflationary observational indices, so the spectral indices of
scalar and tensor perturbations and the tensor-to-scalar ratio.
Thus finding the expression and values of these conformal
invariant quantities in the Einstein frame guarantees that these
have the same values in the Jordan frame.

\section{Essential Features of Higgs Inflation in the Einstein Frame}

As we mentioned in the introduction, having the Higgs field
playing the role of the inflaton is very well motivated
theoretically for many reasons, but mainly for the fact that one
does not need to introduce extra scalar degrees of freedom to
describe the inflaton. The only scalar field that has ever been
observed observationally is the Higgs field, so if this scalar
field is simultaneously the inflaton, this offers an elegant and
theoretical description of the inflationary era and the
electroweak phase transition. However, the large self-coupling of
the Higgs field makes the minimally coupled scalar theory
non-compatible with the CMB observations \cite{Mishra:2018dtg}. If
instead considering the Higgs scalar theory in the minimally
coupled theory, one considers the non-minimally coupled Higgs
theory, this can provide a viable Higgs inflationary theory. In
this section we review in brief the non-minimally coupled Higgs
inflationary theory, which is quite successful in describing the
inflationary era
\cite{Bezrukov:2008ut,Garcia-Bellido:2008ycs,George:2013iia,Mishra:2018dtg}.
The vacuum non-minimally coupled Higgs inflationary action is,
\begin{equation}\label{ta}
\mathcal{S}=\int d^4x\sqrt{-g}\Big{[}f(\phi)R-\frac{1}{2}g^{\mu
\nu}\partial_{\mu}\phi\partial_{\nu}\phi-U(\phi)\Big{]}\, ,
\end{equation}
with the function $f(\phi)$ being the non-minimal coupling and the
Higgs potential $U(\phi)$ being equal to,
\begin{equation}\label{fofphi}
f(\phi)=\frac{1}{2}\left(m^2+\xi \phi^2 \right)\, ,
\end{equation}
\begin{equation}\label{jordanframepothiggs}
U(\phi)=\frac{\lambda}{4} \left(\phi^2-v^2\right)^2\, ,
\end{equation}
with $m^2$ being equal to,
\begin{equation}\label{msquarehiggs}
m^2=M_p^2-\xi v^2\, ,
\end{equation}
$\xi$ being the non-minimal coupling constant whose value is
determined by the constraints on the amplitude of the scalar
perturbations, and it is equal to $\xi=1.62\times 10^4$, and $M_p$
is the reduced Planck mass, and $v$ is the electroweak scale which
in terms of the reduced Planck mass is $v=246\,$GeV$=1.1\times
10^{-16}\,M_p$. Note that potentials similar to the Higgs
potential above appear in curved spacetime
multiplicatively-renormalizable scalar theories \cite{serg2}.
Since the electroweak scale is much smaller than the scale of
inflation $\phi\sim \mathcal{O}(M_p)$, we have that $m\sim M_p$,
thus the non-minimal coupling function becomes,
\begin{equation}\label{nonminimalcoupapprox}
f(\phi)\sim \frac{M_p^2}{2}\left(1+\frac{\xi \phi^2}{M_p^2}
\right)\, .
\end{equation}
Upon performing the following conformal transformation,
\begin{equation}\label{ta1higgs}
\tilde{g}_{\mu \nu}=\Omega^{2}g_{\mu \nu}\, ,
\end{equation}
following the methods of the previous section, we obtain the
Einstein frame canonical Higgs scalar field action, expressed in
terms of the Einstein frame canonical Higgs scalar field
$\varphi$,
\begin{equation}\label{ta5higgs}
\mathcal{S}_E=\int
d^4x\sqrt{-\tilde{g}}\Big{(}\frac{M_p^2}{2}\tilde{R}-\frac{1}{2}
\tilde{g}_{\mu \nu}\partial^{\mu}\varphi
\partial^{\nu}\varphi-V(\varphi)\Big{)}\, ,
\end{equation}
where the ``tilde'' denotes quantities evaluated in the Einstein
frame, and $V(\varphi)$ being equal to,
\begin{equation}\label{einsteinframescalarpotential}
V(\varphi)=\frac{U(\phi(\varphi))}{\Omega^4}\, ,
\end{equation}
and also we have,
\begin{equation}\label{conformalreltionsclarfield}
\frac{d \varphi}{d
\phi}=\frac{1}{\Omega^2}\sqrt{\Omega^2+\frac{6\xi^2\phi^2}{M_p^2}}\,
.
\end{equation}
For $\phi\gg \frac{M_p}{\sqrt{\xi}}$, we have,
\begin{equation}\label{sclarfieldvraphi}
\varphi=\sqrt{6}M_p\ln \left(\frac{\sqrt{\xi \phi}}{M_p}\right)\,
,
\end{equation}
and for $\varphi\gg \sqrt{\frac{2}{3}}\frac{M_p}{\xi}$, the Higgs
potential in the Einstein frame is approximated as follows,
\begin{equation}\label{Higgspotentialeinsteinframeapprox}
V(\varphi)\simeq
V_0\left(1-\exp\left(-\sqrt{\frac{2}{3}\frac{\varphi}{M_p}}
\right) \right)^{-2}\, ,
\end{equation}
where $V_0$ is given by,
\begin{equation}\label{vodefinitoon}
V_0=\frac{\lambda M_p^4}{4\xi^2}\, .
\end{equation}
The amplitude of the scalar perturbations is,
\begin{equation}\label{amplitudeofscalarpert}
\Delta_s^2=\frac{1}{24\pi^2}\frac{V(\phi_i)}{M_p^4}\frac{1}{\epsilon(\phi_i)}\,
,
\end{equation}
and according to the Planck data we must have
$\Delta_s^2=2.2\times 10^{-9}$, so $V_0\sim 9.6\times
10^{-11}\,M_p$. The resulting Higgs inflation phenomenology in the
Einstein frame is viable since it yields a scalar spectral index
of primordial perturbations $n_s=0.967$ and a tensor to scalar
ratio $r=0.003$ which puts the model in the sweet-spot of
compatibility in the Planck data. Furthermore, the tensor spectral
index for the Higgs inflation model is $n_{\mathcal{T}}=-r/8$,
which will be used later on for the calculation of the energy
spectrum of the gravitational waves, in which we will assume a
Higgs inflationary era. Also since the scalar and tensor spectral
indices and the tensor-to-scalar ratio are conformal invariant
quantities, this means that the Jordan frame theory has the same
values for these quantities.

\section{Reheating in Higgs Inflation and the Reheating Temperature}

During the Higgs inflationary era, the temperature dropped
gradually until the end of inflationary era dropped to zero. Thus
it is possible that the electroweak symmetry in the SM broke
during the inflationary era. This is an interesting issue since it
aligns with the phase transition during inflation perspective
already pointed out in the literature
\cite{Jiang:2015qor,Hashida:1998is,Baccigalupi:1997re}, however we
just mention this important issue. After inflation ends, the Higgs
is at its minimum oscillating and its couplings to the SM
particles will reheat the Universe. The question is how large will
the reheating temperature be. During this reheating, if the
reheating temperature becomes larger than $150\,$GeV, the
electroweak symmetry will be restored and the it will break again
later on in the radiation domination era when the temperature
drops again to approximately $150\,$GeV. The evaluation of the
reheating temperature in Einstein frame scalar field theories is
more or less standard
\cite{Amin:2014eta,Choi:2016eif,Dimopoulos:2019gpz,Fei:2017fub,Martin:2014nya,Gong:2015qha,Cai:2015soa,Cook:2015vqa,Rehagen:2015zma,deFreitas:2015xxa,deHaro:2016hsh,Ueno:2016dim,Eshaghi:2016kne,Maity:2016uyn,Tambalo:2016eqr,deHaro:2017nui,Oikonomou:2017bjx,Maity:2017thw,Artymowski:2017pua,German:2018wrx,Ji:2019gfy,Freese:2017ace,Fonseca:2018xzp,Dimopoulos:2018wfg},
so by using the Einstein frame Higgs potential, we shall have an
estimation of the Higgs reheating temperature. We must however
note that in some approaches, the Higgs field itself can affect
drastically the reheating era, by blocking it or even delaying it
\cite{Freese:2017ace, Fonseca:2018xzp} however we shall not
consider such effects here, since we need an estimate of the Higgs
reheating temperature to be used in the calculation of the
gravitational wave energy spectrum.
\begin{figure}[h!] 
\centering
\includegraphics[width=35pc]{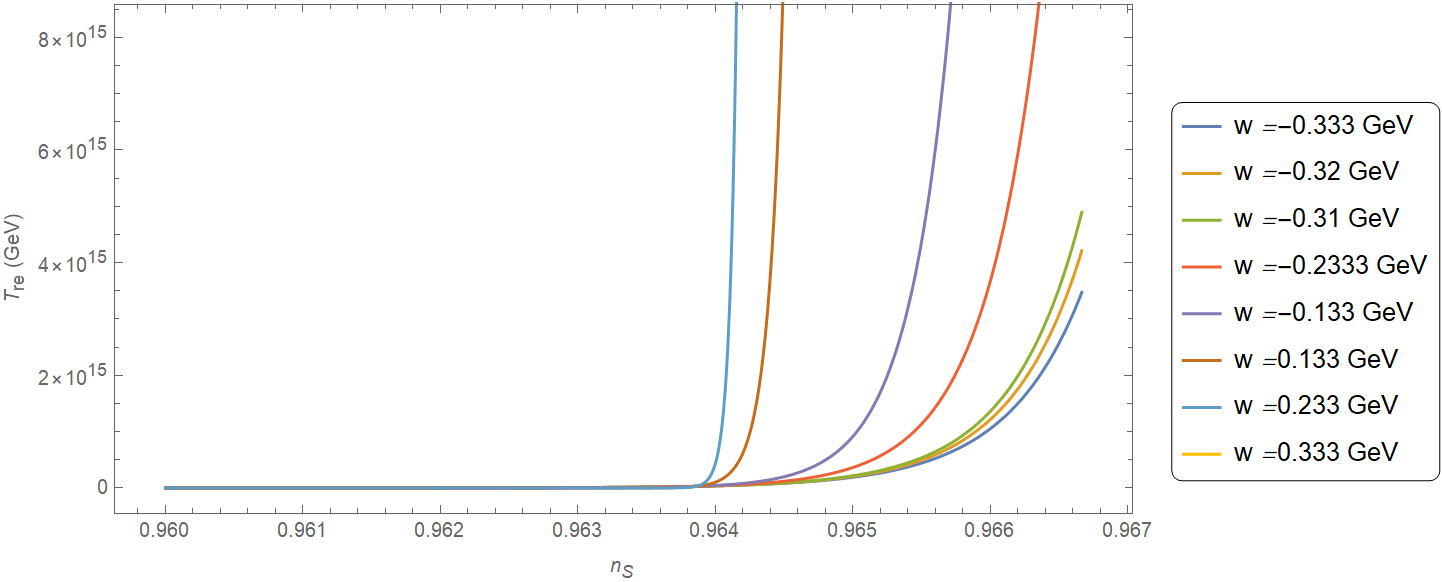}
\caption{The reheating temperature $T_{re}$ (GeV) versus the
scalar spectral index for the Higgs inflation for various values
of the scalar spectral index (depending on the total duration of
the inflationary era which we took to be $50-60$ $e$-foldings) and
for various values of the reheating background EoS parameter $w$.
} \label{fig:TreNs1}
\end{figure}
In order to cover all the possible scenarios though, in the
gravitational waves section we shall also assume a low-reheating
temperature, which however be taken to be larger than the
temperature for which the electroweak phase transition occurs.
Thus the aim in this section is to evaluate the reheating
temperature $T_{re}$ of the Universe, and the total duration of
the reheating era, measured by the reheating $e$-foldings number
$N_{re}$. A question remains regarding the background EoS of the
Universe during the reheating era, which can be in the range
$-1/3<w<1$. So in the following we define $N_{re}$ to be the
duration of the reheating era and specifically the $e$-foldings
number of the reheating era starting when the $e$-foldings number
of the inflationary era is $N_{fin} \sim 50-60$ until the
background EoS of the Universe reaches the value $w =
\frac{1}{3}$, at which point the temperature of the Universe is
$T_{re}$. See Ref. \cite{Cook:2015vqa} for an interesting
discussion on this topic. Following the notation of Refs.
\cite{Cook:2015vqa}, $a_{fin}$ and $\rho_{fin}$ denote the scale
factor and the energy density at the end of the inflationary era,
and also $a_{re}$ and $\rho_{re}$ are the same quantities at the
end of the reheating era, so by using the Friedmann equation we
have,
\begin{equation}
\dfrac{\rho_{fin}}{\rho_{re}} = \left( \dfrac{a_{fin}}{a_{re}}
\right)^{-3(1+w_{re})} \, ,
\end{equation}
with $w_{re}$ being the barotropic index at the end of the
reheating era. Using this, we have \cite{Cook:2015vqa},
\begin{equation}\label{eq:Duration1}
N_{re} = \dfrac{1}{3(1+w_{re})} \ln \left(
\dfrac{\rho_{fin}}{\rho_{re}} \right) = \dfrac{1}{3(1+w_{re})} \ln
\left( \dfrac{45 V_{fin}}{\pi^2 g_{re} T_{re}^4} \right) \, ,
\end{equation}
due to the fact that $\rho_{fin} = \dfrac{3}{2} V_{fin}$ and
$\rho_{re} = \dfrac{\pi^2}{30} g_{re} T_{re}^4$, with $V_{fin}$
being the value of the Higgs potential at the end of the
inflationary era, and also and $g_{re}$ denotes the number of
relativistic degrees of freedom in the reheating era which is
$\sim 100$. Furthermore, the reheating temperature is
\cite{Cook:2015vqa},
\begin{equation}
T_{re} = T_{0} \left( \dfrac{a_{0}}{a_{re}} \right) \left(
\dfrac{43}{11 g_{re}} \right)^{\frac{1}{3}} = T_{0} \left(
\dfrac{a_{0}}{a_{eq}} \right) e^{N_{RD}} \left( \dfrac{43}{11
g_{re}} \right)^{\frac{1}{3}} \, ,
\end{equation}
with $a_{eq}$ being the scale factor at exactly the
matter-radiation equality time instance and in addition $N_{RD}$
is the $e$-foldings number of the radiation-dominated era, since
we have $a_{re} = a_{eq} e^{N_{RD}}$, and also $T_0$ denotes the
present day temperature $T_0\sim 3.19\times 10^{-5}$eV. Also we
have $T_{re}$ \cite{Cook:2015vqa},
\begin{equation}\label{eq:Temperature1}
T_{re} = a_{0} T_{0} \dfrac{H_{k}}{k} e^{-N} e^{-N_{re}} \left(
\dfrac{43}{11 g_{re}} \right)^{\frac{1}{3}} \, ,
\end{equation}
with $N_{k}$ and $H_{k}$ being the duration of the inflationary
era and the Hubble rate of the pivot scale $k = a_{k} H_{k}$ at
the first horizon crossing. With $w \neq \dfrac{1}{3}$, upon
combining Eqs. (\ref{eq:Duration1}) and (\ref{eq:Temperature1}),
we get,
\begin{equation}\label{eq:Duration2}
N_{re} = \dfrac{4}{1 - 3 w_{re}} \left( 61.6 - N_{k} - \ln \left(
\dfrac{V_{fin}^{\frac{1}{4}}}{H_{k}} \right) \right) \, ,
\end{equation}
thus the reheating temperature is \cite{Cook:2015vqa},
\begin{equation}\label{eq:Temperature2}
T_{re} = \left( \dfrac{43}{11 g_{re}}
\right)^{\frac{1}{3}}\dfrac{T_{0} H_{k}}{0.05}
e^{-N_{k}}\,e^{-N_{re}}\, .
\end{equation}
For the scalar field we have,
\begin{equation*}
N_{k} = \int_{t_{k}}^{t_{fin}} H(t) \mathrm{d}t = \kappa
\int_{\phi_{fin}}^{\phi_{k}} \dfrac{V(\phi)}{
\frac{\mathrm{d}V}{\mathrm{d}\phi}} \mathrm{d} \phi \, ,
\end{equation*}
This can easily be evaluated for the Higgs potential, for the
initial and final values of the Higgs scalar at the beginning and
the end of the inflationary era.
\begin{figure}[h!] 
\centering
\includegraphics[width=35pc]{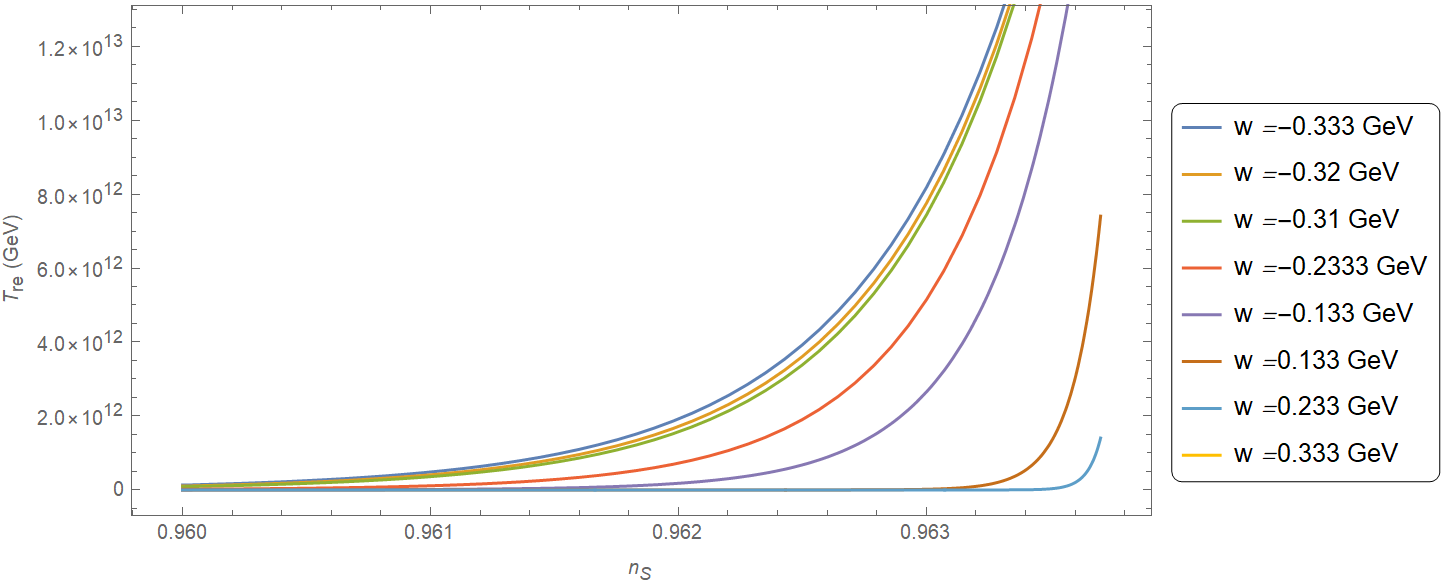}
\caption{The reheating temperature $T_{re}$ (GeV) versus the
scalar spectral index for the Higgs inflation for various values
of the scalar spectral index (depending on the total duration of
the inflationary era which we took to be $50-60$ $e$-foldings) and
for various values of the reheating background EoS parameter $w$.
} \label{fig:TreNs2}
\end{figure}
Using the essential inflationary features of the Higgs potential,
in Figs. \ref{fig:TreNs1} and \ref{fig:TreNs2} we plot the
reheating temperature for the Higgs inflation for various values
of the scalar spectral index (depending on the total duration of
the inflationary era) and for various values of the reheating
background EoS parameter $w$. As it can be seen, the reheating
temperature for the Higgs model varies in the range
$10^{11}-10^{15}\,$GeV, so in the section where we shall study the
energy spectrum of the gravitational waves, the values for the
reheating temperature that belong in this range will be the most
important. However, for completeness, we shall also use lower
reheating temperatures, varying from $500\,$GeV to $10^7\,$GeV.

\section{Electroweak Phase Transition: Second Order, First Order or Crossover Phase Transition?}

\begin{figure}
\centering
\includegraphics[width=35pc]{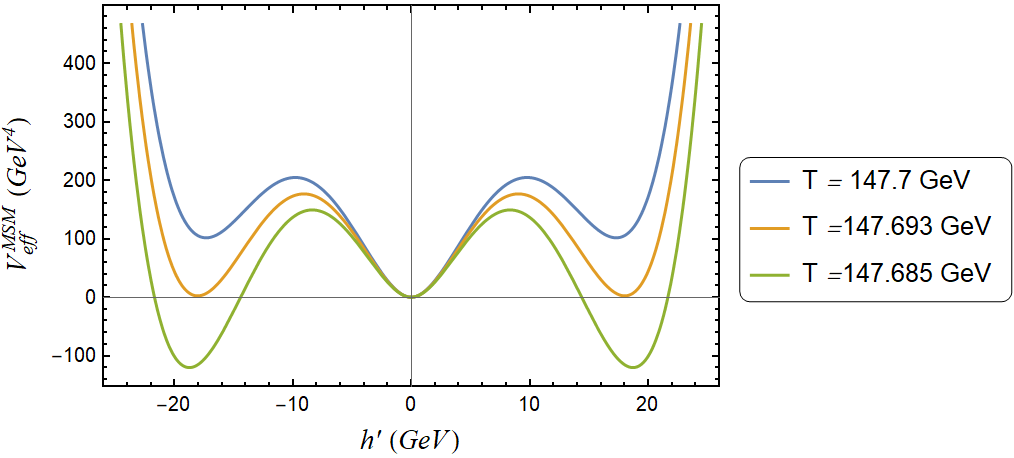}
\caption{The effective potential of the SM for various
temperatures near the critical temperature $T_c\sim 147.693\,$GeV.
The phase transition is a very weak first order transition, so
weak that it qualifies for a crossover or second order phase
transition.}\label{plot1}
\end{figure}

The electroweak symmetry breaking is a controversial phase
transition which was believed to be first order for quite some
time \cite{Anderson:1991zb,
Quiros:1999jp,Arnold:1992rz,Carrington:1991hz}, a feature
depending on the Higgs mass, although numerical studies revealed
that the electroweak phase transition was second order
\cite{Kajantie:1996qd,Kajantie:1996mn}. A low mass Higgs, of the
order $\sim 60-70\,$GeV would favor the first order phase
transition perspective however the Higgs has a mass $\sim
125\,$GeV, thus the electroweak phase transition is a crossover
phase transition. In this section we shall study the
finite-temperature effective potential of the SM using daisy graph
corrections for the thermal masses at all orders. We shall examine
the behavior of the effective potential, for the Yukawas chosen as
in the SM and also we shall reveal the behavior of the origin near
the critical point at which the phase transition occurs. The
1-loop effective potential of the SM is equal to
\cite{Anderson:1991zb,
Quiros:1999jp,Arnold:1992rz,Carrington:1991hz,Morrissey:2012db,Dine:1992wr,Dolan:1973qd,Senaha:2020mop},
\begin{equation}\label{eq:A1}
\begin{split}
    V^{MSM}_{eff} (h, T) = & - \frac{\mu^{2}_{H}}{2} {h}^2 + \frac{\lambda_{H}}{4} {h}^4 + \sum_{i} (-1)^{F_i} n_i \frac{m^4_{i}(h)}{64 \pi^2}\left[ \ln \left( \frac{m^2_{i}(h)}{\mu^2_R}\right) - C_i \right] - \frac{n_t m^4_{t}(h)}{64 \pi^2}\left[ \ln \left( \frac{m^2_{t}(h)}{\mu^2_R}\right) - C_t \right] \\
    & + \sum_{i} \frac{n_iT^4}{2 \pi^2} J_{B} \left(\frac{m^2_i (h)}{T^2}\right) - \frac{n_t T^4}{2 \pi^2} J_{F} \left(\frac{m^2_t (h)}{T^2}\right) \\
    & + \sum_{i} \frac{\overline{n}_i T}{12\pi} \left[m^3_i(h') - \left(M^2_i(h,T) \right)^{3/2} \right],
\end{split}
\end{equation}
and we took into account the Higgs, Goldstone bosons, gauge bosons
and the heaviest quark contribution. In Eq. (\ref{eq:A1}), we
denoted
 \(i = \{h, \chi, W, Z, \gamma \}\)
the Higgs, the Goldstone bosons and the gauge bosons which we took
into account. The daisy corrected thermal expansion of the Higgs
effective potential reads,
\begin{equation}\label{eq:A2}
\begin{split}
    V^{MSM}_{eff} (h, T) = & - \frac{\mu^{2}_{H}}{2} {h}^2 + \frac{\lambda_{H}}{4} {h}^4+ \frac{m^2_{h} (h)}{24}{T}^2 - \frac{T}{12 \pi} \left[m^2_{h} (h) + \Pi_{h} (T)\right]^{3/2} + \frac{m^4_{h}(h)}{64 \pi^2} \left[\ln \left(\frac{a_b {T}^2}{\mu^2_R}\right) -\frac{3}{2} \right] \\
    & + \frac{3 m^2_{\chi} (h)}{24}{T}^2 - \frac{3T}{12 \pi} \left[ m^2_{\chi} (h) + \Pi_{\chi}(T) \right]^{3/2} + \frac{ 3 m^4_{\chi} (h)}{64 \pi^2} \left[\ln \left(\frac{a_b {T}^2}{\mu^2_R}\right) -\frac{3}{2} \right] \\
    & + \frac{6 m^2_{W} (h)}{24}{T}^2 - \frac{4T}{12 \pi} m^3_{W} (h) - \frac{2T}{12 \pi}\left[ m^2_{W} (h) + \Pi_{W_L} (T) \right]^{3/2} + \frac{ 6 m^4_{W} (h)}{64 \pi^2} \left[\ln \left(\frac{a_b {T}^2}{\mu^2_R}\right) -\frac{5}{6} \right] \\
    & + \frac{3 m^2_{Z} (h)}{24}{T}^2 - \frac{2T}{12 \pi} m^3_{Z} (h) - \frac{T}{12 \pi} \left[  M^2_{Z_{L}} (h, T) \right]^{3/2} + \frac{ 3 m^4_{Z} (h)}{64 \pi^2} \left[\ln \left(\frac{a_b {T}^2}{\mu^2_R}\right) -\frac{5}{6} \right] \\
    & + \frac{12 m^2_{t} (h)}{48}{T}^2 - \frac{ 12 m^4_{t} (h)}{64 \pi^2} \left[\ln \left(\frac{a_f {T}^2}{\mu^2_R}\right) -\frac{3}{2} \right] - \frac{T}{12 \pi}\left[ M^2_{\gamma_{L}} (h, T) \right]^{3/2},
\end{split}
\end{equation}
where $a_b=223.0993$ and $a_f=13.943$ and in addition the exact
definitions of the field-dependent masses are given below,
\begin{equation}\label{eq:A3}
    m^2_{h} (h) = - \mu^2_{H} + 3\lambda_{H} {h}^2,
\end{equation}
\begin{equation}
    m^2_{\chi'} (h) = - \mu^2_{H}+ \lambda_{H} {h}^2,
\end{equation}
and furthermore,
\begin{equation}\label{effectivemassW}
    m^2_{W} (h) = \frac{{g}^2}{4} {h}^2,
\end{equation}
\begin{equation}\label{effectivemassZ}
    m^2_{Z} (h) = \frac{{g}^2 + {\tilde{g}}^{2}}{4} {h}^2,
\end{equation}
\begin{equation}\label{effectivemasstop}
    m^2_{t} (h) = \frac{y^2_{t}}{2} {h}^2,
\end{equation}
where \(g\),\(\tilde{g}\) and \({y}_{t}\) stand for the
\(SU(2)_L\), \(U(1)_Y\) and also the top quark Yukawa couplings of
the SM. In Table \ref{}
\begin{table}[h!]
\centering
\begin{tabularx}{0.5\textwidth} {
  | >{\centering\arraybackslash}X
  | >{\centering\arraybackslash}X
  | >{\centering\arraybackslash}X
   | }
 \hline
Particle & Mass (GeV) \\
 \hline
 $h$ & $m_{h}=125\,$GeV &  \\
 \hline
 $W$ & $m_{W}=80.4\,$GeV & $g=0.653$ \\
 \hline
 $Z$ & $m_{Z}=91.2\,$GeV & $\tilde{g}=0.349$ \\
 \hline
 $t$ & $m_{t}=173\,$GeV & $y_t=0.993$ \\
 \hline
\end{tabularx}
\caption{SM particles Yukawas and Masses} \label{table1}
\end{table}
Also the mass parameter $\mu_{H}$ in the Higgs sector is
$\mu_{H}=88.3883\,$GeV so the $\sim {h}^4$ Higgs self-coupling
$\lambda_{H}$ is equal to $\lambda_{H}=0.128$. Furthermore, the
temperature-dependent self-energy corrections of the Goldstone
bosons and of the Higgs particle are equal to,
\begin{equation}\label{eq:A4}
    \Pi_{h} (T) = \Pi_{\chi} (T) = \left(\frac{3{g}^2 }{16} + \frac{{\tilde{g}}^{2}}{16}  +\frac{y^2_{t} }{4} + \frac{\lambda_{H}}{2}\right) {T}^2
\end{equation}
and also the thermal masses of the gauge bosons are equal to,
\begin{equation}\label{T-W}
    \Pi_{W_L} (T) = \frac{11}{6}{g}^2 {T}^2,
\end{equation}
\begin{equation}\label{Z-thermalmass}
    M^2_{Z_L} = \frac{1}{2} \left[ \frac{1}{4} \left({g}^2 + {\tilde{g}}^{2}\right) {h}^2 + \frac{11}{6} \left({g}^2 +{\tilde{g}}^{2} \right) {T}^2 + \sqrt{\left({g}^2 - {\tilde{g}}^{2} \right)^2 \left( \frac{1}{4} {h}^2 + \frac{11}{6}  {T}^2 \right)^2  + \frac{{g}^2 {\tilde{g}}^{2}}{4} {h}^4 }\right],
\end{equation}
\begin{equation}\label{Photon-thermalmass}
    M^2_{\gamma_L} = \frac{1}{2} \left[ \frac{1}{4} \left({g}^2 + {\tilde{g}}^{2}\right) {h}^2 + \frac{11}{6} \left({g}^2 + {\tilde{g}}^{2} \right) {T}^2 - \sqrt{\left({g}^2 - {\tilde{g}}^{2} \right)^2 \left( \frac{1}{4} {h}^2 + \frac{11}{6}  {T}^2 \right)^2  + \frac{{g}^2 {\tilde{g}}^{2}}{4} {h}^4 }\right].
\end{equation}
For the thermal phase transition study of the SM which follows, we
shall take the renormalization scale to be $\mu_R = 2\,m_{t'}$. In
Fig. \ref{plot1} we present the behavior of the Higgs effective
potential at various temperatures near the critical temperature
$T_c\sim 147.693\,$GeV. The phase transition seems to be a typical
case of a first order phase transition, at least structurally. But
the reality seems quite different, since the phase transition is
very very weak first order, which qualifies it to be a crossover
phase transition, or even a second order phase transition. Let us
elaborate further on this.
The strength of a first order phase transition can be revealed by
using the sphaleron rate criterion,
\begin{equation}\label{sphaleron_rate}
    \frac{\upsilon_c}{T_c} > 0.6 - 1.4.
\end{equation}
In the case of the SM, the fraction $\frac{\upsilon_c}{T_c}$ is of
the order $\frac{\upsilon_c}{T_c}\sim \mathcal{O}(0.121)$. Hence,
the phase transition is very weak first order, and thus it is
highly questionable whether bubble nucleation and vacuum
penetration will proceed formally following the first order phase
transition pattern. Hence, the phase transition will likely be a
crossover or a second order phase transition. This is the line of
research we shall adopt in this work, and it is known that in such
second order phase transitions, or even at marginally crossover
ones, the phase transition proceeds with the scalar field rolling
classically from the minimum at the origin towards its new minimum
\cite{Athron:2023xlk}. Of course, such a physical picture would
leave many questions and problems unanswered, like for example the
explanation of the baryon asymmetry in our Universe, which is
generated by a departure from thermal equilibrium guaranteed by a
first order phase transition. This is why a hidden singlet sector
is added in the SM which is coupled only to the Higgs, in order to
render the phase transition a strong one, and indeed such a
physical scenario strengthens the first order phase transition,
see for example \cite{Oikonomou:2024jms}. However, in this work we
shall assume that the Higgs is not coupled to some hidden scalar,
thus the electroweak phase transition proceeds as a second order
phase transition, with the Higgs classically rolling towards its
new minimum, with the new minimum being developed at a critical
temperature of the order $\sim 150\,$ GeV. This era is deeply in
the radiation domination era and the temperature $\sim 150\,$ GeV
corresponds to wavenumbers $k\sim 10^6\,$Mpc$^{-1}$. Now, the
classical rolling of the Higgs may occur in a slow-roll of a
fast-roll fashion, and this may affect the background EoS
parameter, shifting it from the radiation domination value
$w=1/3$. This total background EoS deformation can generate a
direct imprint on the gravitational wave background energy
spectrum. This is the study of the next section.

\section{Second Order Electroweak Phase Transition and Imprint on the Primordial Gravitational Waves}

\begin{figure}[h!]
\centering
\includegraphics[width=40pc]{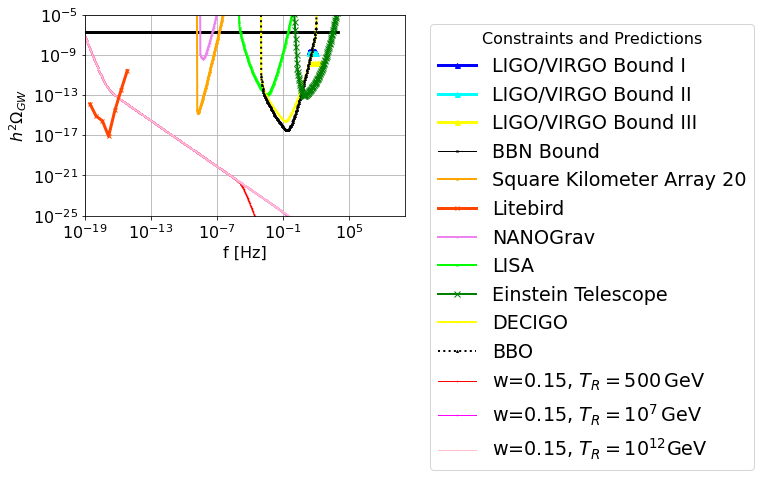}
\caption{The $h^2$-scaled gravitational wave energy spectrum for
the Higgs inflation model with a Higgs second order phase
transition,. The deformed background EoS has the value $w=0.15$
and we considered $T_R=500\,$GeV, $T_R=10^7\,$GeV, and
$T_R=10^{12}\,$GeV. }\label{plot3}
\end{figure}

Assuming a second order electroweak phase transition for the
Higgs, in this section we shall seek the impact of this second
order phase transition on the energy spectrum of the stochastic
gravitational waves.  The second order phase transition occurs at
a temperature $T\sim 150\,$GeV, so this will affect tensor modes
that reenter the horizon at this epoch, with wavenumber $k\sim
10^{6}\,$Mpc$^{-1}$. The classical roll of the Higgs field might
occur in a slow-roll or a fast-roll approach. Our analysis
indicated that the fast-roll will not have any direct effect on
the energy spectrum of the stochastic gravitational waves, so we
focus on the slow-roll Higgs scenario. With a slow-roll Higgs, the
Higgs EoS will be $w_H\sim -1$, thus the total EoS will be smaller
than the radiation domination era value $w=1/3$. We shall consider
two values of the deformed total EoS parameter for our analysis,
specifically $w=0.25$ and $w=0.15$. Regarding the inflationary
era, the Higgs inflation prediction is $n_{\mathcal{T}}=-r/8$ for
the tensor spectral index, and also the tensor-to-scalar ratio is
$r=0.003$. Also the reheating temperature is predicted to be
$10^{11}-10^{15}\,$GeV, but for completeness we shall consider
three reheating temperatures, one low value specifically,
$T_R=500\,$GeV, an intermediate reheating temperature
$T_R=10^7\,$GeV, and a high reheating temperature scenario with
$T_R=10^{12}\,$GeV.
\begin{figure}[h!]
\centering
\includegraphics[width=40pc]{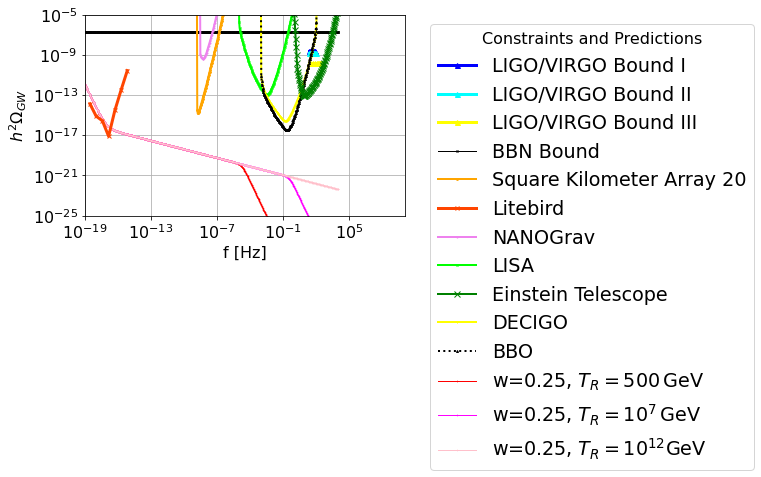}
\caption{The $h^2$-scaled gravitational wave energy spectrum for
the Higgs inflation model with a Higgs second order phase
transition,. The deformed background EoS has the value $w=0.25$
and we considered $T_R=500\,$GeV, $T_R=10^7\,$GeV, and
$T_R=10^{12}\,$GeV. }\label{plot2}
\end{figure}
The slow-rolling of the Higgs field during the radiation
domination era has a direct effect on the total EoS parameter
deforming it, and this has an imprint on the stochastic
gravitational waves. Specifically, the deformation of the EoS
results in a multiplication factor of the form $\sim
\left(\frac{k}{k_{s}}\right)^{r_c}$, with $r_c=-2\left(\frac{1-3
w}{1+3 w}\right)$ \cite{Gouttenoire:2021jhk}, where  $k_{s}$ is
the wavenumber at the moment when the EoS deformation occurs, at a
wavenumber $k_s\sim 10^{6}$Mpc$^{-1}$. Therefore, the $h^2$-scaled
energy spectrum of the stochastic gravitational waves for the
Higgs second order phase transition becomes,
\begin{equation}\label{GWspecfRnewaxiondecay}
\Omega_{\rm gw}(f)=S_k(f)\times
\frac{k^2}{12H_0^2}r\mathcal{P}_{\zeta}(k_{ref})\left(\frac{k}{k_{ref}}
\right)^{n_{\mathcal{T}}} \left ( \frac{\Omega_m}{\Omega_\Lambda}
\right )^2
    \left ( \frac{g_*(T_{\rm in})}{g_{*0}} \right )
    \left ( \frac{g_{*s0}}{g_{*s}(T_{\rm in})} \right )^{4/3} \nonumber  \left (\overline{ \frac{3j_1(k\tau_0)}{k\tau_0} } \right )^2
    T_1^2\left ( x_{\rm eq} \right )
    T_2^2\left ( x_R \right )\, ,
\end{equation}
with $S_k(f)$,
\begin{equation}\label{multiplicationfactor1}
S_k(f)=\left(\frac{k}{k_{s}}\right)^{r_s}\, ,
\end{equation}
and $k_{ref}$ is the CMB pivot scale
$k_{ref}=0.002$$\,$Mpc$^{-1}$. Also, $T_{\rm in}$ denotes the
horizon reentry temperature,
\begin{equation}
    T_{\rm in}\simeq 5.8\times 10^6~{\rm GeV}
    \left ( \frac{g_{*s}(T_{\rm in})}{106.75} \right )^{-1/6}
    \left ( \frac{k}{10^{14}~{\rm Mpc^{-1}}} \right )\, ,
\end{equation}
and also the transfer function $T_1(x_{\rm eq})$ is,
\begin{equation}
    T_1^2(x_{\rm eq})=
    \left [1+1.57x_{\rm eq} + 3.42x_{\rm eq}^2 \right ], \label{T1}
\end{equation}
where $x_{\rm eq}=k/k_{\rm eq}$ and $k_{\rm eq}\equiv a(t_{\rm
eq})H(t_{\rm eq}) = 7.1\times 10^{-2} \Omega_m h^2$ Mpc$^{-1}$,
and in addition the transfer function $T_2(x_R)$ is defined,
\begin{equation}\label{transfer2}
 T_2^2\left ( x_R \right )=\left(1-0.22x^{1.5}+0.65x^2
 \right)^{-1}\, ,
\end{equation}
with $x_R=\frac{k}{k_R}$, and also,
\begin{equation}
    k_R\simeq 1.7\times 10^{13}~{\rm Mpc^{-1}}
    \left ( \frac{g_{*s}(T_R)}{106.75} \right )^{1/6}
    \left ( \frac{T_R}{10^6~{\rm GeV}} \right )\, ,  \label{k_R}
\end{equation}
where $T_R$ denotes the reheating temperature. Furthermore,
$g_*(T_{\mathrm{in}}(k))$ is \cite{Kuroyanagi:2014nba},
\begin{align}\label{gstartin}
& g_*(T_{\mathrm{in}}(k))=g_{*0}\left(\frac{A+\tanh \left[-2.5
\log_{10}\left(\frac{k/2\pi}{2.5\times 10^{-12}\mathrm{Hz}}
\right) \right]}{A+1} \right) \left(\frac{B+\tanh \left[-2
\log_{10}\left(\frac{k/2\pi}{6\times 10^{-19}\mathrm{Hz}} \right)
\right]}{B+1} \right)\, ,
\end{align}
where $A$ and $B$ are,
\begin{equation}\label{alphacap}
A=\frac{-1-10.75/g_{*0}}{-1+10.75g_{*0}}\, ,
\end{equation}
\begin{equation}\label{betacap}
B=\frac{-1-g_{max}/10.75}{-1+g_{max}/10.75}\, ,
\end{equation}
with $g_{max}=106.75$ and $g_{*0}=3.36$. Moreover,
$g_{*0}(T_{\mathrm{in}}(k))$ can be evaluated by using Eqs.
(\ref{gstartin}), (\ref{alphacap}) and (\ref{betacap}), by making
the replacement $g_{*0}=3.36$ with $g_{*s}=3.91$. In Figs.
\ref{plot2} and \ref{plot3} we plot the $h^2$-scaled gravitational
wave energy spectrum for a Higgs second order deformed background
EoS with $w=0.25$ (Fig. \ref{plot2}) and $w=0.15$ (Fig.
\ref{plot3}), with the deformation occurring for frequencies
$k_s=10^{0}$Mpc$^{-1}$. We considered three distinct reheating
temperatures, $T_R=500\,$GeV, $T_R=10^7\,$GeV, and
$T_R=10^{12}\,$GeV are used.
As it can be seen, the effect of a deformation of the background
EoS parameter occurring at the epoch when the temperature is of
the order $\sim 150\,$GeV, caused by a Higgs second order
electroweak phase transition, can have a direct and detectable
imprint on the energy spectrum of the primordial gravitational
waves. Specifically the predicted spectrum can be detected by
Litebird and in fact the case $w=0.25$ can be detected for lower
values of the $h^2$-scaled gravitational wave energy spectrum.
This pattern is unique and also it is mentionable that Higgs
inflation alone cannot predict such a detection pattern of tensor
spacetime perturbations.

\section*{Discussion and Concluding Remarks}

In this work we assumed that the Higgs field non-minimally coupled
to gravity is responsible for the inflationary era. After
reviewing the formalism of non-minimally coupled to gravity single
scalar field inflation, we also presented how the Higgs field can
generate a viable inflationary era. Also we demonstrated how large
can the reheating temperature be in the context of Higgs field
inflation. Beyond that, we showed that the electroweak phase
transition in the context of the SM is a very weak first order
phase transition, a fact that motivates us to think that the
actual electroweak phase transition is actually a second order
phase transition. Using this line of argument, we showed that a
second order phase transition during the radiation domination era
would have a direct observational imprint on the energy spectrum
of the primordial gravitational waves, affecting modes with
frequencies equal or smaller to the ones entering the Hubble
horizon when the Higgs second order phase transition occurs. The
reason for this is the fact that when the Higgs field experiences
a second order phase transition, it slowly-rolls toward its new
minimum from the origin, and this slow-rolling has a direct effect
on the total EoS parameter during the radiation domination era. In
fact it deforms it making it smaller that the radiation domination
value $w=1/3$. We evaluated in some detail the effect of the Higgs
second order phase transition on the energy spectrum of the
primordial gravitational waves and we showed that a unique
observability pattern is generated, in which the signal of such a
transition can be detected by the Litebird experiment. What we did
not consider is the effects of curvature and high energy
corrections to the Higgs effective potential during the
inflationary era \cite{Masina:2024ybn}. Such a possibility is of
course feasible, and the curvature corrections should also be
considered for this era in the effective potential. Also possible
modifications of the inflationary era and also the possibility of
having an inflationary phase transition should be carefully
examined. Furthermore, the inclusion of the axion field is a
viable option for the Higgs field, since this would affect the
reheating temperature of the Higgs inflationary model, if for
example the axion is a kinetic axion, in the fashion of
\cite{Oikonomou:2022tux}. We hope to address such issues in a
future work.

\section*{Acknowledgments}

This research has been is funded by the Committee of Science of
the Ministry of Education and Science of the Republic of
Kazakhstan (Grant No. AP26194585) (Vasilis K. Oikonomou).

\end{document}